\documentclass[aip, jcp, amsmath]{revtex4-1}

\usepackage{graphicx}
\usepackage{dcolumn}
\usepackage{bm}
\usepackage{subfigure}

\begin{document}

\title{The Barrier Method: A Technique for Calculating Very Long Transition Times}

\date{\today}

\author{D. A. Adams}
\affiliation{Department of Physics, University of Michigan,  Ann Arbor Michigan, 48109-1040, USA }

\author{L. M. Sander}
\altaffiliation{Center for the Study of Complex Systems, University of Michigan, Ann Arbor Michigan, 48109-1040,  USA}

\affiliation{Department of Physics, University of Michigan,  Ann Arbor Michigan, 48109-1040, USA }

\author{R. M. Ziff}
\altaffiliation{Center for the Study of Complex Systems, University of Michigan, Ann Arbor Michigan, 48109-1040,  USA}

\affiliation{ Department of Chemical Engineering, University of Michigan, Ann Arbor Michigan, 48109-2136,  USA }

\begin{abstract}
In many dynamical systems there is a large separation of time scales between typical events  and ``rare" events which can be the cases of interest. Rare-event rates are quite difficult to compute numerically, but  they are of considerable practical importance in many fields: for example transition times in chemical physics and extinction times in epidemiology can be very long, but are quite important. We present a very fast numerical technique that can be used to find long transition times (very small rates) in low-dimensional systems, even if they lack detailed balance. We illustrate the method for a bistable non-equilibrium system introduced by Maier and Stein and a two-dimensional (in parameter space) epidemiology model.   
\end{abstract}

\maketitle

\section{Introduction}
Important events for a transition may have a time scale  many orders of magnitude larger than typical events; in this case, they are called ``rare events."  They have been studied in a number of different contexts, including the extinction of diseases \cite{andersson2000stochastic} or of populations \cite{bartlett1961stochastic}, network queue overflow \cite{medhi2003stochastic}, and slow chemical reactions \cite{van2007stochastic}. 

\begin{figure}
\begin{center}
{\includegraphics[width=0.48\textwidth]{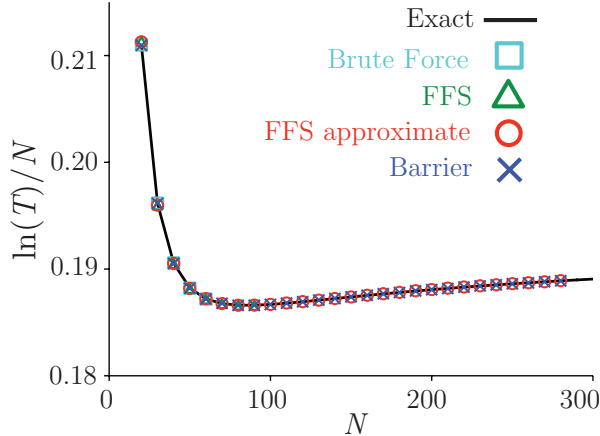}}
\caption{The mean time to extinction, $T$, as a function of the population size, $N$, for the simple SIS model calculated exactly \cite{doering2005extinction} and by two of the numerical methods to be described in the text. All numerical results agree with the exact result within statistical error.}
\label{sis1d}
\end{center}
\end{figure}

To fix our ideas we consider a very simple problem, that of an endemic disease which fluctuates to extinction. Consider a population of fixed size, $N$, with $S$ members who are susceptible to an infection, and $I$ who are infected. When $S$ encounters $I$, the infection is transferred with rate $\beta SI/N$; $\beta$ measures the infectivity.  Infected agents can spontaneously recover with rate $\kappa$, and can be immediately reinfected. This is called the SIS (Susceptible-Infected-Susceptible) model\cite{Jacquez93}. In this simple form it can be thought of as a Markov process in the number of infected; note that $S=N-I$. Thus:
\begin{equation}
\begin{split}
\label{eq:SISoneD}
W(I \to I+1) &=  \beta SI/N \\
W(I \to I-1) &=  \kappa I. 
\end{split}
\end{equation}
There is a long-lived state where the disease persists when $R_0 = \beta/\kappa >1$, namely $I=N(1-R_0^{-1})$. However, there is an important rare event, namely a fluctuation to $I=0$, which means that the disease is extinct and cannot return.  
For this simple form of  the model the mean exit time for this transition, $T$, can be found exactly\cite{doering2005extinction}. For large $N$, there is an asymptotic formula:
\begin{equation}
T \to \frac{R_0}{(R_0-1)^2}\sqrt{\frac{2\pi}{N}} e^{N(\log{R_0}-1+1/R_0)} .
\end{equation}
This formula has features that are generic to the kind of ``barrier climbing" problems that we treat here. The exit time is of the form $g(N)\exp(NW)$ where $g$ is a slowly varying prefactor, and $W$ is a generalized barrier height (or quasipotential) scaled by the large parameter, $N$. A plot of the exact results for the SIS model with $R_0=2$  is given in Figure \ref{sis1d} along with results from numerical computations that we will describe below. Note that even for modest-sized systems the mean time to extinction can be huge: for $N=300$ we have $T \approx 4 \times 10^{24}$.

As we see from this simple case, these phenomena are frequently out of reach for brute-force simulations. To overcome this problem, many techniques have been developed \cite{dellago2008transition}. In this article, we revisit this problem and present a very efficient technique which we call  the \emph{barrier method} and which gives the mean first-passage time for a transition to an unlikely target state. The method does not depend on special features such as knowledge of a steady-state distribution or detailed balance in the process. We need only that the dynamics be stochastic, and that the transition probabilities depend on the current state. That is, we deal with Markov dynamics, which can be reversible or irreversible. 

The essence of the algorithm is to follow the development of an ensemble of systems and oversample the cases that happen to approach the target, and not allow backtracking away from the target. In this respect our method resembles the signposting algorithm \cite{Adams2008Percolation,Adams2009DLA} that we  developed for finding the penetration of a random walker into a fractal. 

We believe that the barrier method is the most efficient method available for computations in relatively low dimensions. In this paper we explain how it differs from previous techniques, and we apply it to two systems. The first is described by  a non-equilibrium model introduced by Maier and Stein.\cite{maier1993effect} The second is a generalization the SIS model in which the population is allowed to fluctuate  \cite{schwartz2009predicting}. In this model, we are concerned with the average time for the disease to go extinct. Lastly, we discuss the advantages of our method and future work.

\section{Background}
To study rare events in a Markov process we must deal with  states in state-space that are unlikely to be visited in any simulation of reasonable length \cite{hammersley1964monte}.   Most sample paths spend the majority of time visiting the most likely states and give good estimates of  the corresponding probability density.  For rarely visited regions, we need to use special methods.

An example of such a method is biased sampling, in which is we arrange our simulation to  be  biased towards rarely visited regions. Two important subsets of this approach are importance sampling and splitting techniques. Importance sampling, the most commonly used method for equilibrium systems, requires some \textit{a priori} information about the probability distribution. In contrast, if the probability distribution is only accessible via simulation,  \emph{splitting} and related techniques are very useful. The latter case is the focus of our work.

\subsection{Splitting and RESTART}
Splitting \cite{hammersley1964monte} involves placing a barrier in state space. When a sample path crosses the barrier it is split into  independent realizations  whose statistical weights add up to the original. By placing one of these splitting barriers in a region which would be infrequently visited, that region will subsequently be better sampled.  For very difficult-to-reach regions of state space one barrier is not sufficient. The use of several barriers is called multilevel splitting \cite{glasserman1999multilevel}. 

Multilevel splitting has two  drawbacks. First, if  the barriers are too close together or too far apart, the number of simulations will grow or decay exponentially.  Further, realizations which have small weight (because they have been split many times) will often \emph{backtrack}, i.e., tend to move back to the well-sampled regions and waste computational effort.  Some of these problems have been solved by RESTART \cite{villen1991restart} (REpetitive Simulation Trials After Reaching Threshold) which is designed for dealing with queuing problems.

In RESTART \cite{villen1991restart} one considers nested subsets of phase space,  $A \supset B \supset C \supset D$, where $D$ is the target. Barriers are placed between these regions.  A sample path is started in  $A$ and evolves until it reaches $B$. Then the sample is split into $R$ `retrials'  with equal weight.  One of these  is designated the primary, and all realizations evolve independently. If any of the non-primary paths backtracks into $A$, it is terminated. Each barrier is crossed in turn, and the time spent  in $D$ by the reweighted paths gives an estimator of the phase space probability in $D$.

RESTART partially solves the backtracking problem because most samples that exit low probability regions are terminated.  However, in the original version \cite{villen1991restart, villen1994restart} it still can lead to a divergence in the number of samples  if the barriers are too closely spaced.   It has been noted that the barrier placement problem could be partially alleviated by performing fixed effort RESTART instead of fixed splitting RESTART \cite{garvels1998comparison}, but this does not completely fix the problem.  We give another approach to this problem below.  

\subsection{Forward Flux Sampling}

Forward flux sampling (FFS)\cite{allen2005sampling} uses the same principles as splitting in applications to computational chemical physics.   Most  rare-event techniques in this area\cite{dellago2008transition}  require equilibrium ensembles and detailed balance. FFS is unusual because it is applicable to systems without detailed balance.   It has been used to study genetic switches \cite{allen2005sampling, allen2006simulating, valeriani2007computing}, nucleation \cite{valeriani2007computing, sanz2007evidence, allen2008homogeneous} and a model problem due to  Maier and Stein\cite{valeriani2007computing, allen2006forward} which we also treat below. 

FFS finds the first passage time  between metastable states $A$ and $D$ as follows. First, we run a single long simulation and count the number of times the sample path exits $A$ through barrier $\lambda_0$, which bounds region $A$. The average flux through the barrier, $k_0$,  is gotten by dividing this number by the total simulation  time, discounting the time associated with trajectories that reach $D$ and return to $A$. 
If we call $\lambda_M$ the barrier around $D$, the transition rate from $A$ to $D$ is:
\begin{equation}
\label{eq:FFS_main}
k_{AD} =k_0 P(\lambda_M | \lambda_0).
\end{equation}
where  $P(\lambda_M | \lambda_0)$ is the probability that a sample path which starts on $\lambda_0$ will cross $\lambda_M$ before going back to $A$.  

We can get  $P(\lambda_M | \lambda_0)$ efficiently by introducing intermediate barriers $\lambda_i, i=1 \dots M-1$ divide the sample space along level surfaces of some reasonable guess for the reaction coordinate -- we call this the order parameter. The probability factors into:
\begin{equation}
\label{eq:FFS_prob}
P(\lambda_M | \lambda_0) = \prod_{i=0}^{M-1} P(\lambda_{i+1} | \lambda_{i}),
\end{equation}
where  $P(\lambda_{i+1} | \lambda_i)$ is the probability of starting at $\lambda_i$ and reaching $\lambda_{i+1}$ before going back to $\lambda_0$.
To measure $P(\lambda_{1} | \lambda_0)$,  $R$ samples are started from the locations along $\lambda_0$ where they left $A$ in the first step. The paths are continued until they reach $\lambda_1$ or  go back inside $\lambda_0$. The fraction of that reach $\lambda_1$ is the estimator of $P(\lambda_1 | \lambda_0)$. Then we proceed to $\lambda_2$ and start $R$ samples, etc.  The point is to break down a long sample path  into a series of short segments.

FFS does not allow the number of samples to diverge, as in splitting. However, it does backtrack because samples which start at $\lambda_i$ must be allowed to return to $A$. This effect can be somewhat reduced by pruning of the backtracking paths; see\cite{allen2006simulating}. However, if there are metastable states the region between $A$ and $D$, backtracking can take a long time. 

Further, the calculation of $k_{0}$ requires that the initial long simulation reaches the end state $D$ at least once in order to properly sample the entire region between the start and the target. If one does this,  FFS will often be  bottlenecked by the calculation of $k_{0}$; this  defeats the purpose of using a rare-event technique.  Fortunately, for systems with featureless barriers,  running an initial simulation that crosses $\lambda_0$ a fixed number of times, say $10R$, is typically sufficient. We call this version ``approximate FFS." For a comparison of the two versions of FFS for the SIS model, see Figures \ref{sis1d}, \ref{fig:Efficiency}. The approximate method of FFS gives good accuracy for this problem, and is quite fast.However, for systems with metastable states in the region between $A$ and $D$, this method of determining $k_{0}$ is not sufficient. Our method (see below) overcomes all of these problems. 

%

\subsection{Milestoning}
Milestoning \cite{faradjian2004computing} is a technique for equilibrium systems in which  one runs simulations of short paths between barriers to find the local first passage time from one barrier to the next. The equilibrium ensemble on the barriers gives the launching points, and there is no backtracking at all. The local first passage times are put into an integral equation to find the global first passage time. As we will see, we avoid using equilibrium considerations on the barriers by keeping track of the landing points of individual paths, but otherwise, our method uses similar ideas. 

\section{Barrier method}

\subsection{Algorithm}

\begin{figure}[t]
\includegraphics[width=0.45\textwidth]{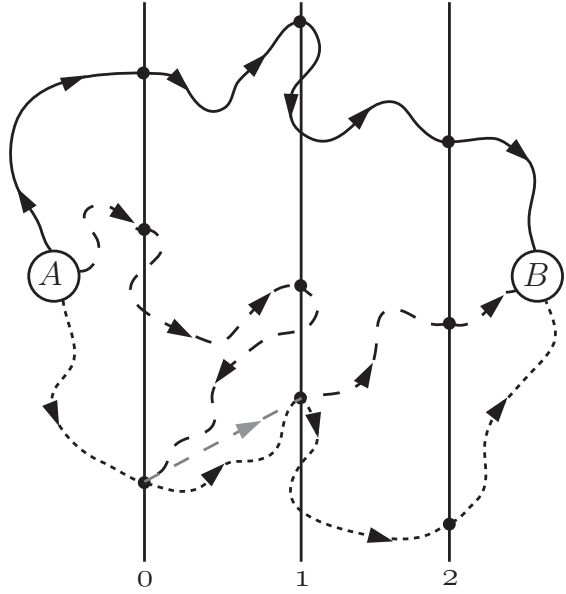}
\caption{\label{fig:Barrier_Fig} Barrier method. Three different paths are started from A. The paths cross barrier $0$ then cross barrier $1$. The middle path then moves backwards and reaches barrier $0$. The end points on barriers $0$ and $1$ are used to ``jump'' the path back to the barrier $1$, grey path. Paths terminate once they reach $B$.}
\end{figure}

In the barrier method, we consider the same sort of problem as for FFS.  For the moment, we assume that the locations of the barriers are known \textit{a priori}, as shown in Fig.~\ref{fig:Barrier_Fig}.  We will discuss the best way to distribute the barriers below. In the first step, $R$ trial simulations are started in $A$ and run until they reach $\lambda_0$.  Each trial $r$ ends at $W_{\lambda_0}^r$ at time $\tau_{A,\lambda_0}^r$. $R$ more trials are started at the locations along $\lambda_0$ where each $r$ ended in the previous step and run until they reach $\lambda_1$. These trials can backtrack as far as they need to.  The locations along $\lambda_1$ where the trials stopped are $W_{\lambda_1}^r$ and the transition times from $\lambda_0$ to $\lambda_1$ are $\tau_{\lambda_0, \lambda_1}^r$.  For each $r$ the total time to start from $A$ and reach $\lambda_1$ is $\tau_{A,\lambda_1}^r = \tau_{A,\lambda_0}^r + \tau_{\lambda_0,\lambda_1}^r$. 

On the next step we eliminate backtracking. We start each trial $r$ at the location on $\lambda_1$ where the previous trial $r$ finished, $W_{\lambda_1}^r$. Each sample path continues until it either reaches $\lambda_{2}$ and stops or returns to $\lambda_{0}$. If the trial returns to $\lambda_{0}$, \emph{we move it back to $\lambda_1$ and add a sample of the time to return from $\lambda_0$ to $\lambda_1$}.  In practice, we find the the closest $W_{\lambda_{0}}^s$ to the current location and add $\tau_{\lambda_{0},\lambda_{1}}^s$ to the trial time and continue the trial at $W_{\lambda_1}^s$. That is, we move the sample point ``in one step" to the next barrier. Continuing,  the sample path can either reach $\lambda_{1}$ and stop or go back to $\lambda_{0}$, in which case we repeat the process of jumping. The locations along $\lambda_{2}$ and the transition time for each $r$ are $W_{\lambda_{2}}^r$ and $\tau_{\lambda_{1},\lambda{2}}^r$. The total time to reach $\lambda_{2}$ is $\tau_{A,\lambda_{2}}^r = \tau_{A,\lambda_{1}}^r+\tau_{\lambda_1,\lambda_{2}}^r$. Then we repeat the process for the next barrier, and continue until $\tau_{A,\lambda_{M}}^r$ and $W_{\lambda_M}$ are calculated. The average transition time from $A$ to $\lambda_M$ is  $ \sum_r \tau_{A,\lambda_M}^r/R$. 

This method differs from FFS in two ways. First we work with transition times instead of transition rates and never deal with probabilities directly. 
Second, the barrier method does not require sample paths to travel from $\lambda_i$ all the way back to $\lambda_0$. This can lead to a dramatic improvement in efficiency over FFS, as we will see in the next section. Note that metastable states in the barrier region pose no problem for this method. 

\subsection{Accuracy and Efficiency}
\begin{figure}[t]
\subfigure[]{\includegraphics[width=0.5\textwidth]{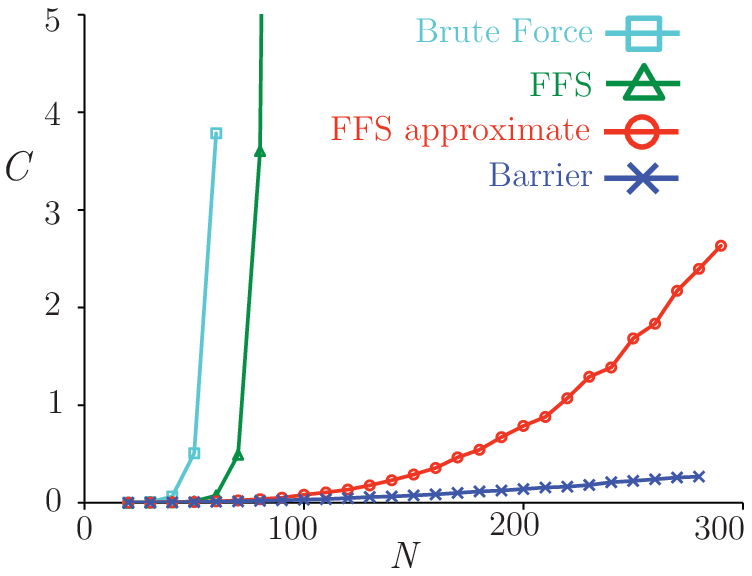}}
\subfigure[]{\includegraphics[width=0.48\textwidth]{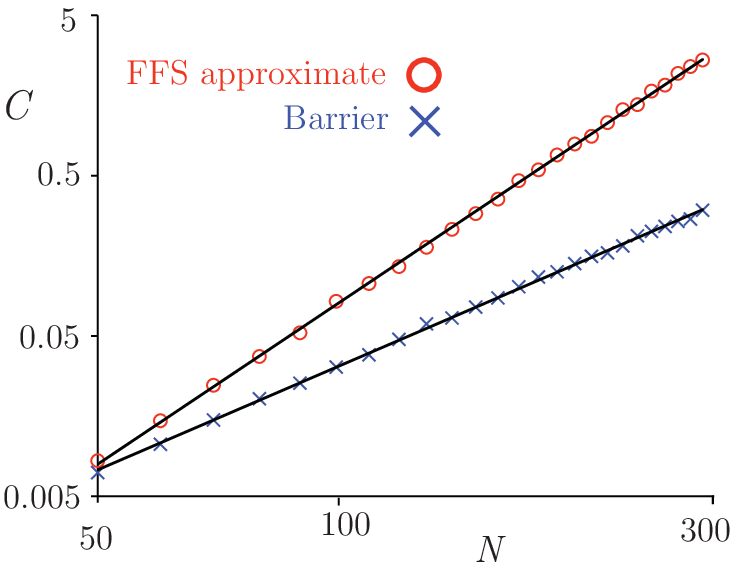}}
\caption{\label{fig:Efficiency}a.) CPU time required to calculate $T$ to a fixed precision (arbitrary units) for the SIS model. For FFS and the barrier method, the same number of barriers were used and placed in the same locations (every 5 steps in number of infected). b) Efficiency comparison between the (approximate) FFS method and the barrier method. The solid lines represent the best fit to a power-law: with powers $3.3$ (FFS) and $2.1$ (barrier method). These values should be taken as rough estimates of the powers. }
\end{figure}

The simple SIS model described above  is exactly soluble. This allows us to make a direct comparison between FFS and the barrier method. Using  $N = 20$ to $200$, we  found less than $1$\% difference between the barrier method  and the exact results, which was within the variance of the measurement. Both versions of  FFS gave similar accuracy; see Fig. \ref{sis1d}. The exact version of FFS, which samples the whole region between $A$ and $D$ in the estimate of $k_0$ is impractical for $N > 60$, as shown in Fig.~\ref{fig:Efficiency}. 

We now compare the efficiency of approximate FFS and the barrier method. Simulation efficiency can be defined by the amount of computation time, $C$, needed to obtain the exit time within a relative accuracy of $\sigma$; we chose $\sigma = 0.1$.  Using this definition, we varied  the number of trials, $R$, to find $C$  for $R_0=2$ for a range of $N$, from $50$ to $290$ in steps of $10$ for FFS and the barrier method. For both we used $N/10-1$ barriers, placed every $5$ infectious population size steps starting at $I=5$. No attempt was made to optimize the barrier placement. The comparison is given in Figure~\ref{fig:Efficiency}. The simulations were performed on a 2.8 GHz Intel processor. We find that both algorithms appear to have a power-law relationship between $C$ and $N$, with powers of roughly $3$ and $2$ for FFS and the barrier method respectively.  This shows that for a case where the barrier method is not obviously better, because of the simple  landscape, it still significantly outperforms FFS. 

\begin{figure}[t]
\includegraphics[width=0.45\textwidth]{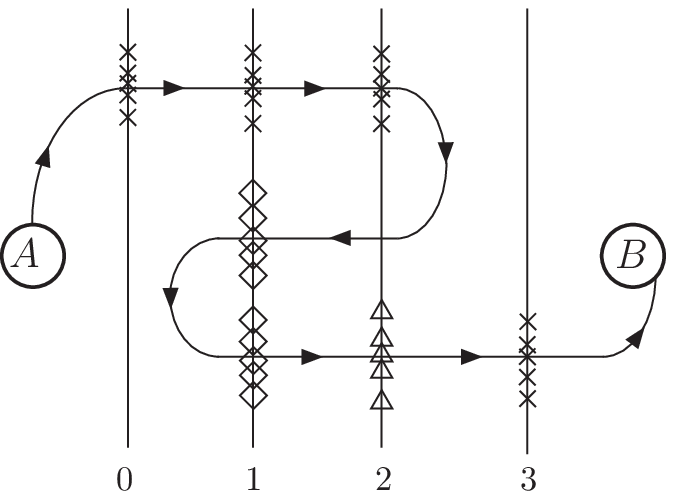}
\caption{\label{fig:Barrier_Backtrack} An example of the barrier method when the most probable exit path (MPEP) crosses barriers several times. The line from $A$ to $B$ represents the MPEP. The $x$'s represent the initial locations where simulations reach a given barrier. These locations are used to generate the original set of lookups. The diamonds represent the new lookups generated by trials originated on the second barrier.  Similarly, the triangles are new lookups generated by the trials started on the third barrier. Note that without adding new lookups, the simulation would never finish because the closest lookups on previous barriers would move the simulations backwards along the MPEP when they are moved to the next barrier.}
\end{figure}

\subsection{Dynamic Barrier Placement}
We have also developed a way to determine the locations of all of the $\lambda_i$ barriers as the simulation progresses. The method outlined in this section can be applied to FFS as well as any other biased sampling methods that use barriers, e.g. RESTART \cite{villen1991restart}. We have previously used a similar method for signposting \cite{Adams2008Percolation,Adams2009DLA}. 

Assuming that the location of the first barrier has been chosen and all of the $W_{\lambda_0}^r$ values have been calculated, $p$ probe trials are started along $\lambda_0$ at $W_{\lambda_0}^r$ for some randomly chosen $r$'s. The probe trials are run for some fixed number of simulation steps, $S$.  The location of $\lambda_1$ is chosen as at the average longest excursion along the order parameter of the probe trials. Next, $W^r_{\lambda_1}$ and $\tau^r_{\lambda_0,\lambda_1}$ are recorded. We repeat the process by running more probe trials for $S$ simulation steps, not counting any steps involved with jumping from $\lambda_0$ to $\lambda_1$, starting at random $W_{\lambda_i}^r$'s.  The average furthest excursion is used as the location of $\lambda_2$. This process is repeated until $\lambda_M$ is reached. An alternative approach would be estimate the location of the next barrier such that a fixed fraction of samples reach $\lambda_{i+1}$ before reaching $\lambda_{i-1}$ when starting at $\lambda_i$. An rough estimate of the optimal forward fraction is\cite{villen1994restart} $e^{-2}$. We did not chose this approach because we found that for small noise the probability of making any forward progress is significantly smaller than $e^{-2}$. This would cause the barrier method to stop making progress. 

There has been some previous work on barrier placement or staging using FFS\cite{Borrero08}. The scheme uses two FFS calculations: one with a guess of the best barrier locations and fixed $R$, and a second with either optimized barrier locations or optimized the number of trials $R$ for each barrier.  The authors found that it is better to optimize the spacing of the barrier to get uniform $p(\lambda_{i+1} | \lambda_{i})$ with fixed $R$ than to optimize $R$ for each barrier. When we apply dynamic barrier placement to FFS, we also take this view. The important difference from the approach outlined above is that we determine the location of the barriers \textit{one at a time} instead of all at once. This  leads to computational gains when the trial barrier placement is significantly different from the optimal placement.

\subsection{Additional lookups}

As with FFS, the barriers are should be set up at fixed values of the order parameter. The closer the order parameter is to the true reaction coordinate, the more efficient the method becomes. For the barrier method, an additional issue arises if the most probable exit path (MPEP) crosses a barrier more than once, so that it goes backwards. The original set of lookups move simulations from a previous barrier to the current barrier \textit{along the paths already discovered}. These lookups cannot move the simulations along the correct MPEP, as seen in Fig.~\ref{fig:Barrier_Backtrack}. This problem can be fixed by allowing additional lookups to be created if needed. For the barrier method applied to the generalized SIS model, below, we used this feature.

\section{Maier-Stein model}

Maier and Stein \cite{maier1993effect} introduced  an interesting example of a dynamic system which lacks detailed balance. It has received considerable theoretical \cite{maier1996scaling, maier1997limiting}, experimental \cite{luchinsky1997experiments, luchinsky1997irreversibility}, and computational interest \cite{luchinsky1997experiments, crooks2001efficient, allen2006forward, valeriani2007computing}. We study two aspects of this model: the mean exit time, $T$, from one of the metastable states and the distribution of exit locations along the separatrix. 

The model is specified by two coupled stochastic differential equations: 
\begin{equation}
\begin{split}
\label{eq:MS}
\dot{x} &= f_x( {\bm {\mathrm x}}) + \xi_x(t), \\
\dot{y} &= f_y( {\bm {\mathrm x}}) + \xi_y(t),
\end{split}
\end{equation}
where $ {\bm {\mathrm x}} = (x,y)$ and ${\bm {\mathrm f}} = (f_x,f_y)$ is the time-independent drift field:
\begin{equation}
\begin{split}
\label{eq:force}
f_x &= x - x^3 - \alpha x y^2, \\
f_y &= -\mu y(1 + x^2).
\end{split}
\end{equation}
For $\alpha =\mu$ the model obeys detailed balance.
The white noise  ${\bm {\mathrm \xi}} = (\xi_x, \xi_y)$ has variance $\epsilon$:
\begin{equation}
\label{eq:noise}
\langle \xi_i(t)\rangle = 0, \quad \langle \xi_i(t+\tau) \xi_j(t) \rangle = \epsilon \delta_{ij} \delta(t-\tau).
\end{equation}
We are interested in the small noise case: $\epsilon \to 0$.
The model is bistable with the metastable states located at ${\bm {\mathrm x}} = (\pm 1, 0)$. There is a separatrix at $x=0$. The exit time, the transition time from one of the metastable states to the separatrix, is, in the small noise limit: 
\begin{equation}
\label{eq:escape}
T \propto  e^{W / \epsilon},
\end{equation}
where $W$ is the generalized barrier height. In this model, $W$ is a function of $\alpha$ and $\mu$. If $\mu=1$, for $1 < \alpha < 4$, there is a unique MPEP, and for $\alpha > 4$ there are two MPEPs.   

\subsection{Simulation}
To simulate the system, we solve Eqs. (\ref{eq:MS}) using the Euler method:
\begin{equation}
\begin{split}
\label{eq:num-int}
x(t+h) &= x(t) + h \left( x(t) - x(t)^3 - \alpha x(t)y(t)^2 \right) 
+  \sqrt{\epsilon h}, \\
y(t+h) &= y(t) + h \left( -\mu y(t)(1+x(t)^2) \right) + \sqrt{\epsilon h}, 
\end{split}
\end{equation}
where $h$ is the time step. 

\subsection{Exit time and distribution of exit points}
We start $R$ simulations at the left metastable state $(-1,0)$ and iterate Eq. (\ref{eq:num-int}) to obtain the trajectories. We dynamically locate each barrier. The first barrier is located at the average furthest excursion along $x$, the order parameter, using $S/4$ probe steps. All subsequent barriers are located after $S$ steps. To calculate the exit time, we have three parameters at our control: the number of trials, $R$, the time step, $h$, and the spacing of the barriers, controlled by $S$. To find $W$, we record the exit time for various values of $\epsilon$ and find the slope of of $1/\epsilon$ vs. $\ln T$.

\begin{figure}[t]
\includegraphics[width=0.48\textwidth]{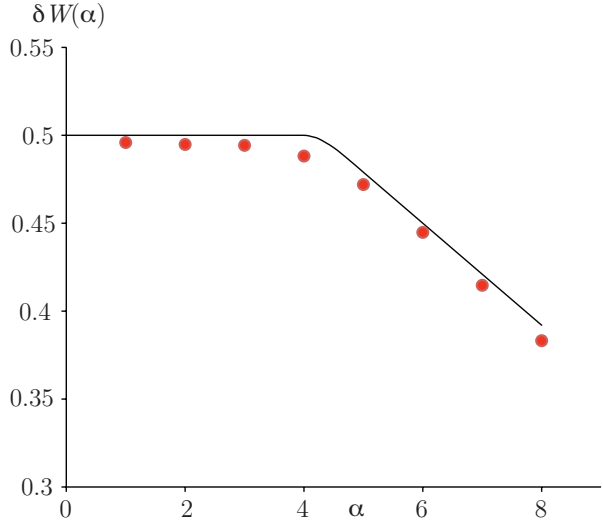}
\caption{\label{fig:Walpha} Barrier height, $W$, for the Maier-Stein model as a function of $\alpha$ for $\mu=1$. Circles, barrier method;  line, analytic theory\cite{maier1993effect} for $\epsilon \to 0$.}
\end{figure}

\subsection{Results}
Using the barrier method we measured $W$ and the exit distributions along the separatrix, $P(y)$. We computed  $W$  for $\mu=1$ for  $\alpha = 1$ to $\alpha = 8$, and compared to analytic theory\cite{maier1993effect}. This results are shown in Fig. \ref{fig:Walpha}. 

In order to use this model, we have to choose  $h$, $R$, and $S$. These parameters have different, competing effects. Increasing $h$ increases the efficiency but decreases the accuracy. The opposite is true for $S$.  Also, effects of the values of $h$ and $S$ are connected; small $S$ causes there to be a large number of barriers, which requires a small $h$ to give an accurate result, and vice versa. We used $h=10^{-6}$, $R=7\cdot 10^4$, and $S = 4 \cdot 10^6$. However, we used a value of $S$ which is four times smaller than the general case to locate the first barrier.  The exit time was measured with five independent trials for five values of $1/\epsilon$: $20$, $40$, $60$, $80$, and $100$, for each $\alpha$. The calculated value of $W$ and the  small-noise theory\cite{maier1993effect} are shown in Fig.~\ref{fig:Walpha}. The numerical results are consistently smaller than the theory and increase as $1/\epsilon$ increases. The values of $W$ in Fig.~\ref{fig:Walpha} were consistent over other values of $h$, $R$, and $S$, that we tested. 

\begin{figure}[t]
\includegraphics[width=0.48\textwidth]{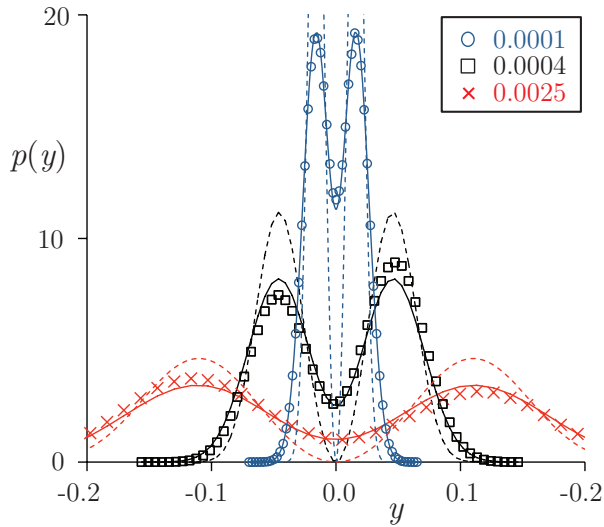}
\caption{\label{fig:ExitDist} The exit distribution for the Maier-Stein model for $\alpha=10, \mu=0.67$ and $\epsilon =0.0025, 0.0004, 0.0001$. The simulation results are denoted by symbols. The dashed line is the  symmetrized Weibull distribution $P(y) = N |y|^{2/\mu-1} \exp{(-|y/A|^{2/\mu}/\epsilon)}$ from theory\cite{maier1997limiting, luchinsky1999observation}, where $N$ is the normalization, $A$ is a parameter of order unity. Solid line: Weibull distribution convolved with a Gaussian with $\sigma = B \sqrt{\epsilon}$.}
\end{figure}

We ran separate simulations to measure the exit distribution along the separatrix. Because we are interested only in the final location along the separatrix and not the time it takes to reach it, we found that we could use significantly larger $h$. We were able to get reliable results for values of $h$ as large as $0.0001$. This allowed us to reach much smaller values of $\epsilon$; see Fig.~\ref{fig:ExitDist}. The parameter values used to obtain the results in Fig.~\ref{fig:ExitDist} are: $h = 0.0001$, $R = 10^5$, and $S = 4\cdot 10^4$. The results were averaged over twenty independent simulations for each $\epsilon$. The significant result is that for the smallest noise value, $\epsilon=0.0001$, the value of $P(0)$ is not close to zero as the theory suggests. Rather the ratio of $P(0)$ to the maximum of $P(y)$  appears instead to be \textit{increasing} as $\epsilon$ decreases. 

These results are consistent with other simulation \cite{crooks2001efficient} and experimental \cite{luchinsky1999observation} results.  Previously, the results were assumed not to match the theory because the values of $\epsilon$  were not small enough. The barrier method allows us to  reach a value of $\epsilon$ which is $50$ times smaller than the best previous simulation result \cite{crooks2001efficient} and $110$ times smaller than the best experimental result \cite{luchinsky1999observation} without $P(0)/P_{max}$ getting any closer to zero. 

We propose that the reason for this is that  the theoretical prediction of the Weibull distribution represents the leading term in $\epsilon$. For finite $\epsilon$ the distribution should be ``rounded" over a scale  $y = O(\sqrt{\epsilon})$\cite{luchinsky1999observation}. Accordingly, we convolved  the asymptotic  theory, the Weibull distribution given in the caption of Fig.~\ref{fig:ExitDist}, with a Gaussian with $\sigma = B \sqrt{\epsilon}$; see Fig.~\ref{fig:ExitDist}. We find good agreement between the simulation results and the convolved theory. The values of $B$ which gave the best fit were $0.8$, $0.85$, and $0.85$ for $\epsilon = 0.0025$, $0.0004$, and $0.0001$, respectively. The fact that $B$ is roughly constant over a wide range of $\epsilon$ gives support to our estimate.

The reason that  $P(0)/P_{max}$ does not tend to  zero is that even though the rounding is over a scale that decreases as $\sqrt{\epsilon}$, the location of the maxima of $P(y)$ also moves toward the origin. These locations are
$y_{max} = \pm 2^{-\mu/2} A \epsilon^{\mu/2}(2-\mu)^{\mu/2}$, 
so that $|y_{max}| \propto A \epsilon^{\mu/2} \approx A \epsilon^{1/3}$ for $\mu = 0.67$. If $A$ were constant, $P(0)/P(y_{max})$ should approach zero rather slowly. We numerically found this rate to be $\epsilon^{0.6}$. However, we find that $A$ is not constant. Our best fit values for $A$, which is unaffected by the convolution, are $0.94$, $0.72$, and $0.4$ for $\epsilon = 0.0025$, $0.0004$, and $0.0001$, respectively, so that $P(0)/P(y_{max})$ does not approach 0 in our computations.

\section{Generalized SIS model}
It is interesting to generalize the SIS model of (\ref{eq:SISoneD}) to allow fluctuations of the total population by introducing birth and death rates\cite{Jacquez93,Dykman08,schwartz2009predicting,Khasin09}. Now there are two independent stochastic variables, $S$, the number of susceptibles, and  $I$, the number of  infected, and four parameters: $\mu$, the birth and death rate assumed equal, $\beta$, the infectious contact rate, $\kappa$, the disease-recovery rate, and $N$, the steady-state population size. The transition rates are now:
\begin{eqnarray}
\label{eq:SISmaster}
W[(S,I) \to (S+1,I)] &=& \mu N, \quad W[(S,I) \to (S-1,I)] = \mu S,  \quad W[(S,I) \to (S,I-1)] = \mu I, \nonumber\\
W[(S,I) \to (S+1,I-1)] &=& \kappa I, \quad
W[(S,I) \to (S-1,I+1)] = { \beta S I }/{N}. 
\end{eqnarray}
The model has an endemic state when $R_0 = {\beta}/{(\mu + \kappa)} > 1$. There is one stable fixed point, the endemic state $(S,I)=(N R_0^{-1}, N(1 - R_0^{-1}))$, and an unstable saddle point $(N,0)$, where the disease is extinct.  We seek the transition time from the endemic state to the disease-free state which will be of the form $T \sim \exp(NW)$, as above.

We will be interested in the case of small $\mu$ so that population fluctuations are slow compared to disease dynamics. It might seem that the situation would be very similar to the case $\mu=0$ treated above. However, this is not true\cite{Khasin09}. Population fluctuations make extinction of the disease much easier: the most likely exit path is via a population decrease at fixed $S$ followed by extinction along a path of smaller fixed population, and then an increase of population of susceptible individuals to $S=N$.

\subsection{Simulation}
We simulate the SIS system using standard techniques\cite{Bortz75,gillespie1976general}. 
We define the order parameter as  $-I$; the barriers are added at decreasing values of $I$. Because this system is on a 2d parameter lattice it is much easier to dynamically add new lookups if previously unexplored backward regions are reached or if a given lookup has been used too frequently. 

The method we use to add lookups is as follows. Every time a sample moves backwards and reaches the previous barrier, a new lookup for that site is generated with a probability $p_G = L_G/D(S,I)$, where $L_G$ is a constant which controls the growth rate of the lookups and $D(S,I)$ is the number of lookups at site $(S,I)$; if $p_G$ is greater than $1$ a new lookup is always added. If a new lookup value is needed, the value is produced by  starting a path at $(S,I)$ until it reaches the next barrier, and stops or reaches the previous barrier. There a lookup is used to move back to the current barrier to continue. Note that a new lookup can move back to a previous barrier and cause another new lookup to be generated at \textit{that} previous barrier. This cascading effect can continue until the first barrier is reached. This effect makes programming the algorithm more complex, but the cascading is necessary to produce accurate results when the MPEP crosses some of the barriers several times, as illustrated in Fig.~\ref{fig:Barrier_Backtrack}. 

\begin{figure}[t]
\subfigure[]{\includegraphics[width=0.48\textwidth]{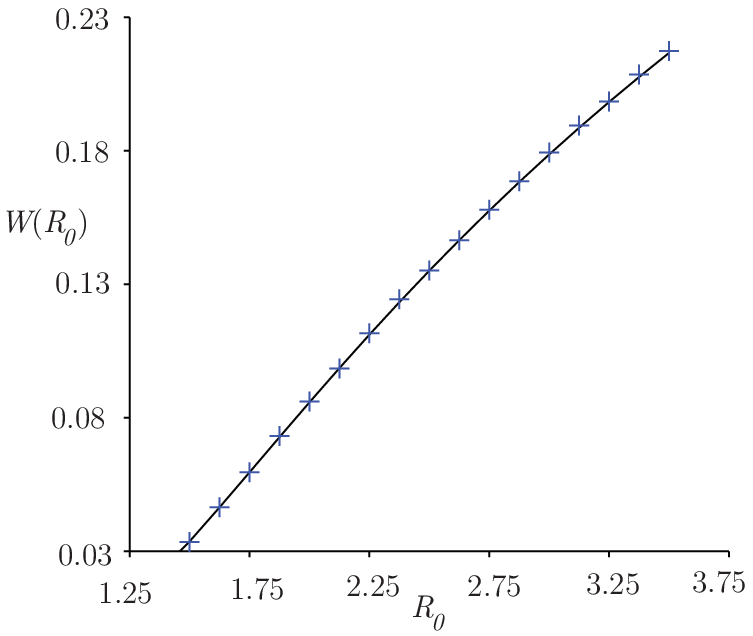}}
\subfigure[]{\includegraphics[width=0.48\textwidth]{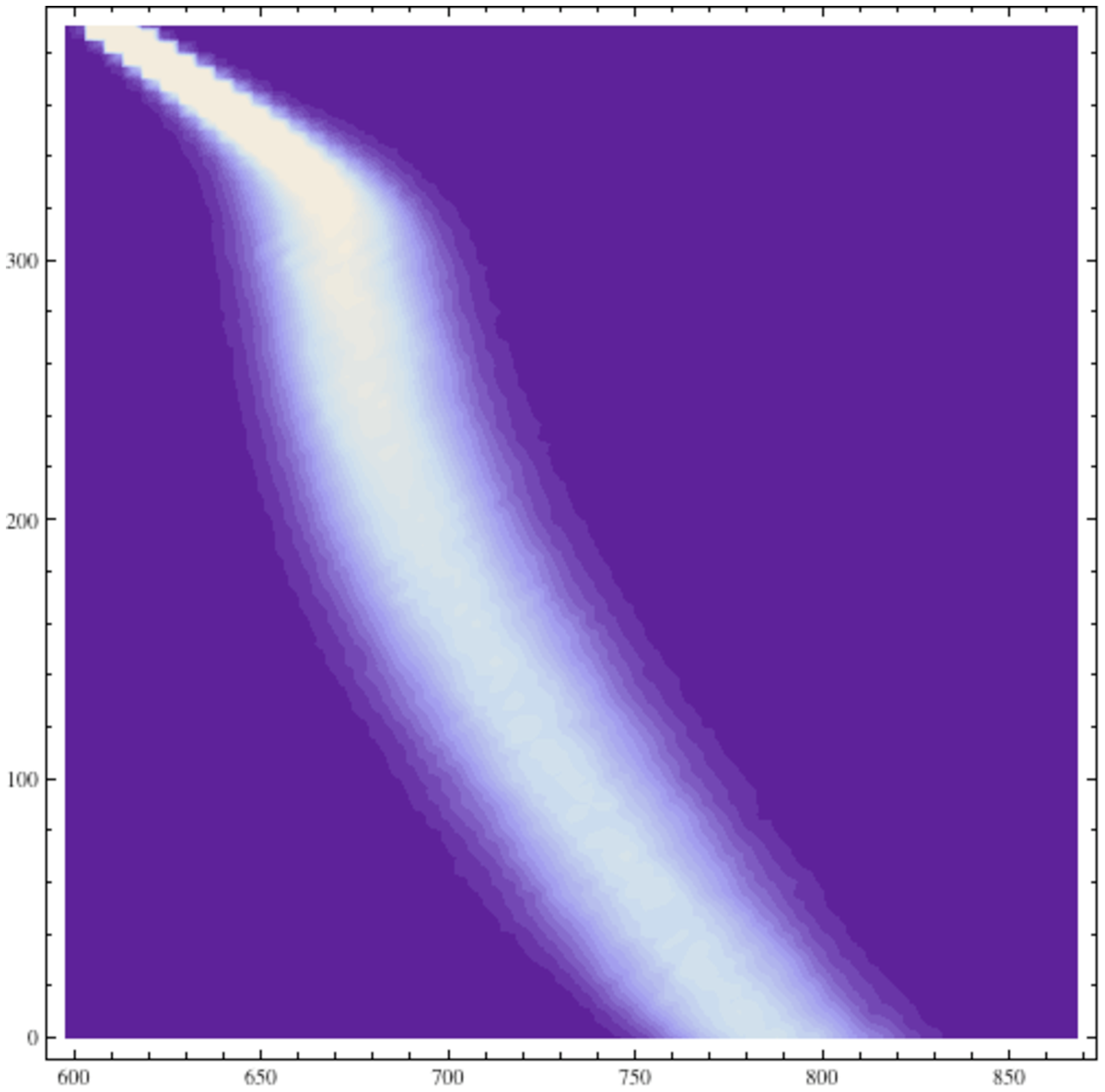}}
\caption{\label{fig:SIS} a.) Generalized barrier in the SIS model for $\mu=1, \kappa=100$ as a function of $R_0$. The line is the theoretical estimate\cite{Khasin09}. b.) Distribution of first passage points on barriers for $N=1000, R_0=1.667, \mu=0.25$. The exit point on the line $I=0$ is in good agreement with the prediction\cite{Khasin09} $S \approx N/\sqrt{R_0}$.}
\end{figure}

\subsection{Results}
\begin{figure}
\includegraphics[width=0.96\textwidth]{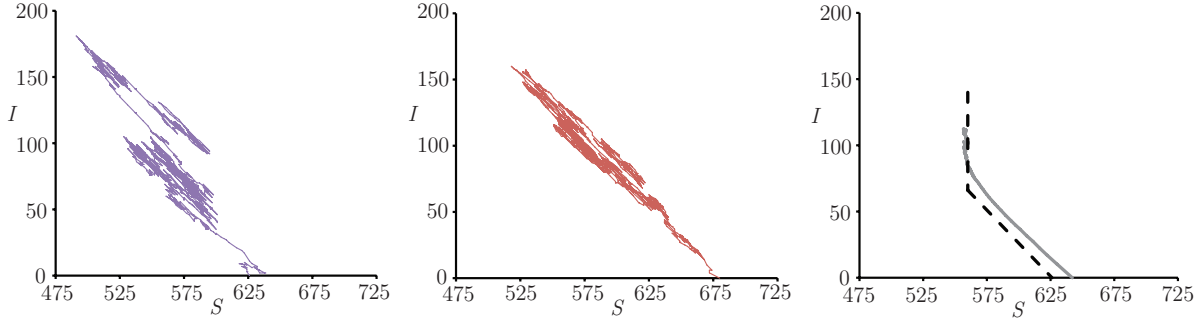}
\caption{\label{exitpath} Exit paths for the SIS model for $N=700, R_0=1.25$. Left panels, some sample paths, right panel, average over many paths to estimate the MPEP with a dashed line for the corresponding theory \cite{Khasin09}. }
\end{figure}

We found the exit time in SIS model, $T \sim \exp(NW)$, for $\mu=1$ and $\kappa=100$ and varied $\beta$ to obtain different values of $R_0$.  We  compare to analytic work \cite{Dykman08,schwartz2009predicting,Khasin09} in Fig.~\ref{fig:SIS} and Fig.~\ref{exitpath}. In these references the authors chose $\mu = 0.02$. They did this because the very large separation in time scales allowed them to use their analytic techniques. 

We calculate $W$ by performing a linear regression of $\ln T$ versus $N$, as in the Maier-Stein section. We ran $10$ simulations for each of following $N$: $50$, $100$, $150$, $200$, $250$, $300$, $350$, $400$, $450$, and $500$. For the last three values of $N$ we did this only for $R_0<2.5$. By looking at the residual of the fit, we found for smaller $R_0$ the smallest sizes were not large enough to reach a constant value of $W$. The smallest $N$ included in the fits were $350$, $350$, $300$, $300$, $250$, $250$, $200$, $200$, and $200$ for $R_0=1.5$, $1.625$, $1.75$, $1.875$, $2.0$, $2.125$, $2.25$, $2.375$, and $2.5$ and greater, respectively. The calculated values of $W$ are shown in Figure~\ref{fig:SIS} and compared to analytic estimates\cite{Khasin09}.

The barrier method does not give the exit path directly. If we plot the distribution of the $W^r$'s on the barriers, it is the distribution of first passage points. However, the paths that continue are not uniformly distributed on the barriers. Nevertheless, it is interesting to plot the first-passage distribution, Fig.~\ref{fig:SIS}.

We did a separate, brute force, computation to find the actual MPEP. A few sample paths are shown in Figure \ref{exitpath} along with the average of many paths.

\section{Discussion}
In this paper, we developed a new rare-event technique, the barrier method. We described the relationship between it and other related techniques, such as FFS, and showed that for a simple model problem  the barrier method is more efficient than FFS, especially for large $N$. 

The barrier method was then used to find exit times and distributions for the Maier-Stein model\cite{maier1993effect}. We found fairly good agreement with theory on exit times and  good agreement with previous simulation and experimental results,  on the exit distributions. We determined that convolving the theory with a Gaussian to account for the next correction to the theory gave excellent agreement for the distribution of exit points.  

The exit times for an SIS model with births and deaths were then calculated. The results agreed with analytic estimates \cite{Dykman08,schwartz2009predicting,Khasin09}. The MPEP was also found for this case.

The barrier method is an excellent tool to determine rare events in low dimensional systems. In this paper we have treated one and two degrees of freedom, and we have preliminary work for three dimensions. We believe the most important aspect of the barrier method is the elimination of practically all `backtracking.' This can allow  traversal of  landscapes with many metastable states .  The method is also general enough to apply to on- and off-lattice problems, equilibrium and non-equilibrium problems and any system that can be written as a non-deterministic Markov process. 

There are limitations to the  method. The most important is that  high-dimensional problems (i.e., systems with many degrees of freedom) are difficult to treat this way.  For example, for nucleation problems in Ising systems of $N$ spins the number of dimensions is $2^N$. The barrier method eliminates backtracking so that each  barrier must be sampled well enough for the lookups  to be accurate, a daunting task in very high dimensions.  Also we need to determine which lookup is closest in a high-dimensional space and then figure out if that lookup is `close enough' or whether another lookup must be created.  

However, the point of view that we take, focussing on times rather than rates, is useful even in high dimensions. We have also developed a variation of the FFS technique which uses this point of view and applied it to a non-equilibrium nucleation problem\cite{Adams2010FFST}. 

\section{Acknowledgments}
This research was supported in part by the National Science Foundation through TeraGrid resources\cite{catlett2007teragrid} provided by Purdue University and through DMS-0553487. We would like to thank C. Doering for useful conversations.

%

\end{document}